\documentclass[12pt]{article}

\usepackage{amsmath,amssymb,epsfig}
\usepackage{graphicx}
\begin{document}
\def\refname{\Large~~~~~~~{\bf References}}
\newcommand{\el}{\left}
\newcommand{\er}{\right}
\newcommand{\p}{\prime}
\newcommand{\rr}{\rho}
\newcommand{\ro}{\rho^\circ}
\newcommand{\ti}{\tilde}
\newcommand{\veps}{\varepsilon}
\newcommand{\dis}{\displaystyle}
\newcommand{\scr}{\scriptsize}

\begin{center}
{\Large \bf Model for Restoration of Heavy-Ion Potentials at
Intermediate Energies
}\\[5mm]
{\large\bf  ~K.M.~Hanna$^{1,2}$, K.V.~Lukyanov$^2$,
V.K.~Lukyanov$^2$, B.~S{\l}owi{\'n}ski$^{3,4}$,
E.V.~Zemlyanaya$^2$}\\[3mm]
{\small\it ~~~~~$^1$Math. and Theor. Phys. Dept., NRC, Atomic Energy
Authority,
Cairo, Egypt}\\
{\small\it $^2$Joint Institute for Nuclear Research, Dubna,
Russia~~~~~~~~~~~~~~~~~~~~~~~~~~~~}\\
{\small\it $^3$Faculty of Physics, Warsaw University of
Technology, Warsaw, Poland ~~~}\\
{\small\it $^4$Institute of Atomic Energy, Otwock-Swierk,
Poland~~~~~~~~~~~~~~~~~~~~~~~~~~~~}\\
\end{center}


\begin{abstract}
{\small Three types of microscopic nucleus-nucleus optical
potentials are constructed using three patterns for their real and
imaginary parts. Two of these patterns are the real $V^H$ and
imaginary $W^H$ parts of the potential which reproduces the
high-energy amplitude of scattering in the microscopic
Glauber-Sitenko theory. Another template $V^{DF}$ is calculated
within the standard double-folding model with the exchange term
included. For either of the three tested potentials, the
contribution of real and imaginary patterns is adjusted by
introducing two fitted factors. An acceptable agreement with the
experimental data on elastic differential cross-sections was
obtained for scattering the $^{16,\,17}$O heavy-ions at about
hundred Mev/nucleon on different target-nuclei. The relativization
effect is also studied and found that, to somewhat, it improves the
agreement with experimental data.}
\end{abstract}

\section {Introduction}
One of the main goals of studying heavy-ion scattering remains to
obtain the nucleus-nucleus optical (complex) potential. Such a
potential is required not only for physical interpretation of
experimental data in elastic channel but also to get the
optical-model wave functions used in the DWBA calculations of
direct inelastic processes and of the nucleons removal reactions.
Unfortunately, when fitting data with the help of phenomenological
optical potentials one cannot obtain their parameters
unambiguously. The other problem is that the parameters of
phenomenological potentials depend on the collision energy, atomic
numbers and isospins of nuclei. These dependencies present many
difficulties in composing appropriate formulae for the global
heavy-ion potentials of scattering.

Therefore one ought to follow the more justified way for searching
the nucleus-nucleus potentials, namely, to develop the respective
microscopic models. In this connection, the attractive and
commonly used models are based on the double-folding (DF)
procedure, where one calculates integrals with overlapping the
density distribution functions of colliding nuclei and the
effective nucleon-nucleon potentials (see, e.g.,
\cite{{SL},{Knyaz},{KS}}). Moreover, the microscopic models arose
considerable interest because they can supply us with underlying
effective NN-forces at normal and higher nuclear densities (see,
e.g., \cite{KSO}). This in-medium dependence of NN-potentials is
of the great importance in both nuclear- and astro-physics where
deeper understanding of e.g. neutron stars and super novae
phenomena is needed.

In nucleus-nucleus scattering at energies near and higher the
Coulomb barrier, most applications were made by using
the optical potential where the real part is calculated within the
microscopic model with the direct and exchange terms included,
while the imaginary part of the potential is taken in a
phenomenological Woods-Saxon form with a three or more adjustable
parameters. In this model, say, semi-microscopic model
\cite{{SL},{Knyaz},{KS}}, one further free parameter is usually
introduced to renormalize the real DF-part of the optical
potential. Thus, the general problem still remains when one
parametrizes the global dependence of the imaginary part on the
potential energy, atomic numbers,... etc.

In the present work, we suggest the method where the pattern potentials
are used to compose the microscopic nucleus-nucleus optical potential.
As a basis we take the complex potential which fully corresponds to
the microscopic high-energy approximation of Glauber and Sitenko
\cite{{Gla},{Sit}}, being later developed in \cite{{Czyz},{Form}} for
deriving the nucleus-nucleus scattering amplitude. This potential
(composed of both the real and imaginary parts)
depends on energy and uses density distributions of nuclei
and the nucleon-nucleon amplitude of scattering with the in-medium effects
included. Besides, we take into consideration the microscopic DF-potential,
the real one, and use it as a pattern for constructing the full
nucleus-nucleus potential. We hope that this regular procedure for obtaining
the complex potentials can protect one against the possible non-physical
forms of phenomenological potentials obtained in the standard fitting
procedure.

In Section 2 the microscopical formulation is presented while Section 3 is
devoted to results, discussions, and some conclusions.

\section {Microscopic Optical Potentials}

To formulate the very complicated many-body scattering problem in terms of
an equivalent optical potential one should to appeal not only to its
theoretical elegance but also to develop the reliable methods which provide
its reasonably simple relation to experimental data. In principle, the
optical potential in its general form as is done, e.g., in \cite{Fesh}, has
a very complicated and nonlocal form. However, one believes that it can be
presented in the equivalent local form by using a realistic localized
expression for the density matrix. So, below we will test the microscopic
nucleus-nucleus energy- and density-dependent optical potential in a compact
form as follows:
\begin{equation}\label{eq1}
U_{opt}(r)\,=\,N_{r} V(r)\,+\,iN_{im}W(r).
\end{equation}
Here the three patterns for both of the real $V(r)$ and imaginary
$W(r)$ parts are calculated by using the appropriate microscopic
models while the normalizing factors $N_{r}$ and $N_{im}$ are
considered as free parameters to be fitted to the experimental
data.

\begin{figure}[t]
\begin{center}
\epsfig{file=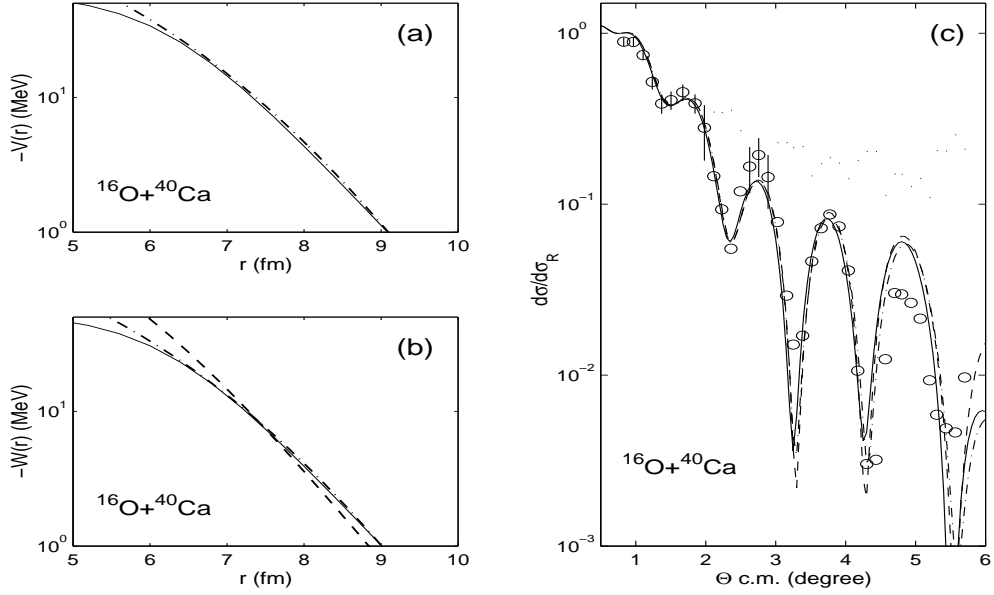,height=8cm,width=.8\linewidth}
\end{center}
\caption{The optical potentials $U_{opt}^B$ and $V_{opt}^C$
obtained basing on the HEA- and DF-patterns with the fitted $N_r$
and $N_{im}$ coefficients (see Table 1A), and the respective
ratios of the elastic differential cross-sections to the
Rutherford one, for $^{16}$O+$^{40}$Ca at $E_{lab}=$1503 MeV.
Panels (a) and (b) are designed for the real and imaginary parts
of potentials, where dashed curves are the real and imaginary
parts of potentials with the patterns $V^H$ (or $W^H$), while
dash-dotted curves are for those with the patterns $V^{DF}$; the
fitted parts of WS-potential from \cite{Ro} are shown by solid
lines. In panel (c), solid curve is calculations with
WS-potential, dashed - with $U_{opt}^B$, and dash-dotted - with
$U_{opt}^C$. Open circles -- experimental data from \cite{Ro}.}
\end{figure}

\begin{figure}[t]
\begin{center}
\epsfig{file=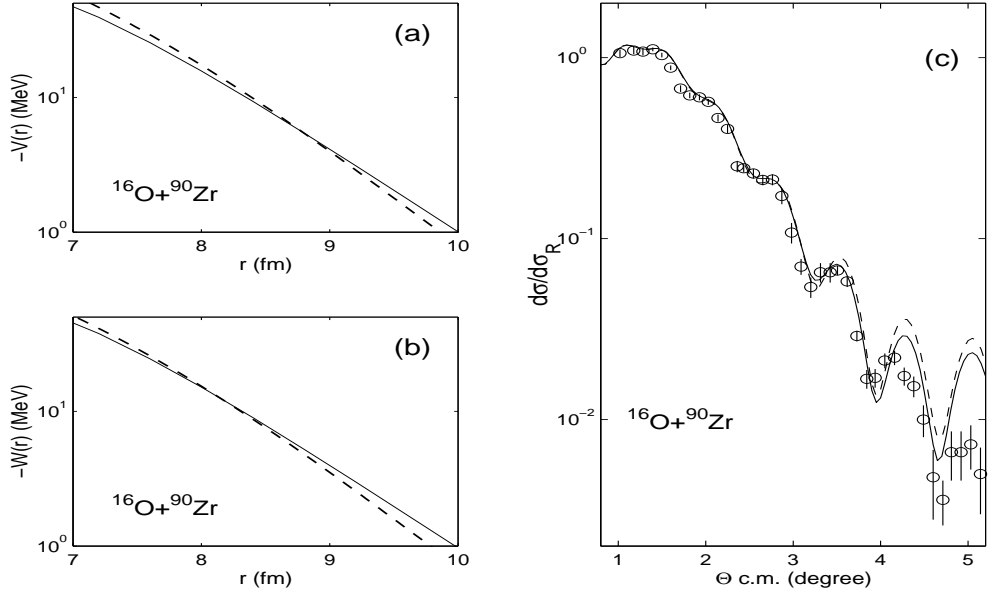,height=8cm,width=.8\linewidth}
\end{center}
\caption{The same as in Fig.1 but for scattering
$^{16}$O+$^{90}$Zr with optical potential $U_{opt}^A$. }
\end{figure}

\begin{figure}[t]
\begin{center}
\epsfig{file=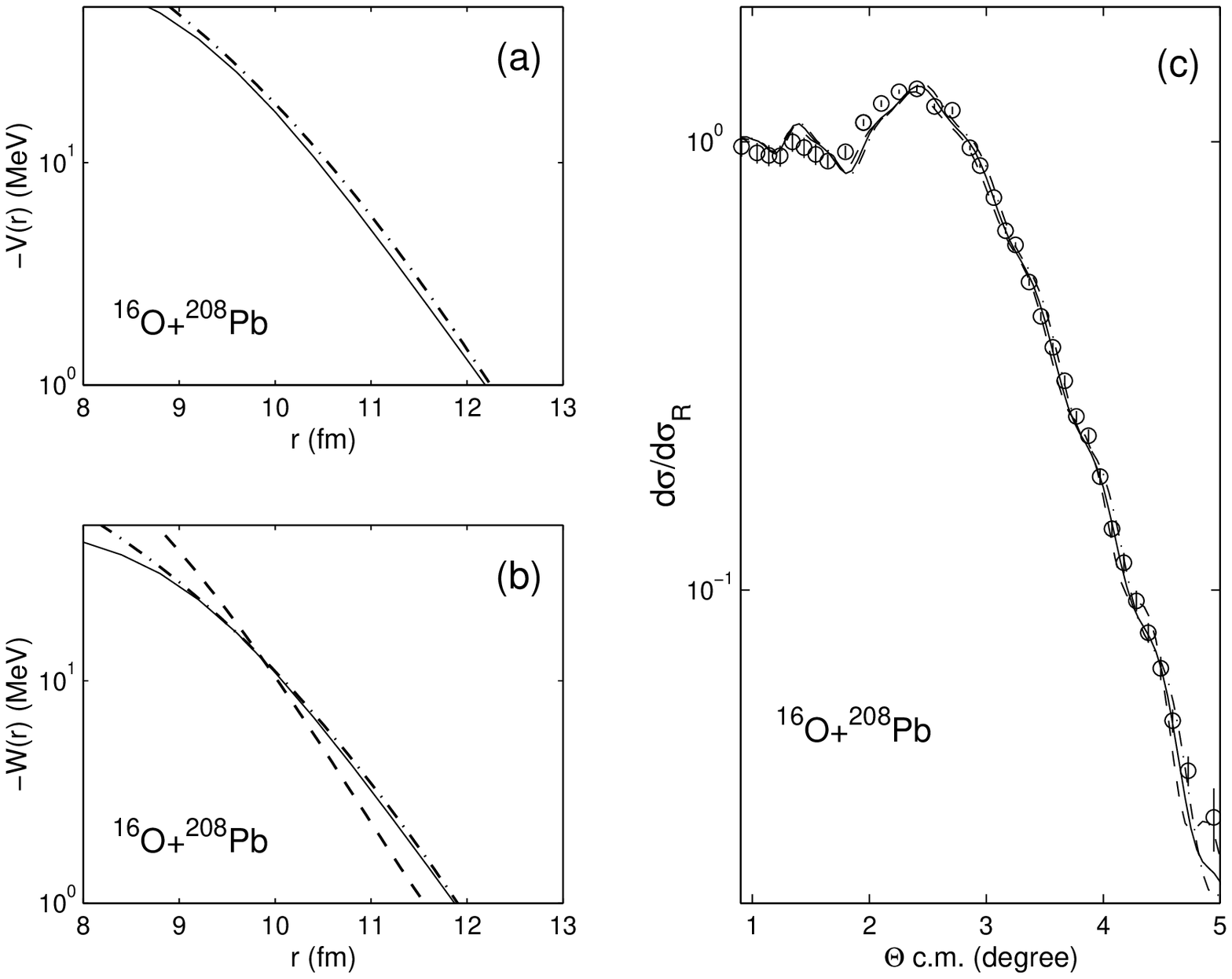,height=8cm,width=.8\linewidth}
\end{center}
\caption{The same as in Fig.1 but for scattering
$^{16}$O+$^{208}$Pb with optical potentials $U_{opt}^B$ and
$U_{opt}^C$.}
\end{figure}

\begin{figure}[t]
\begin{center}
\epsfig{file=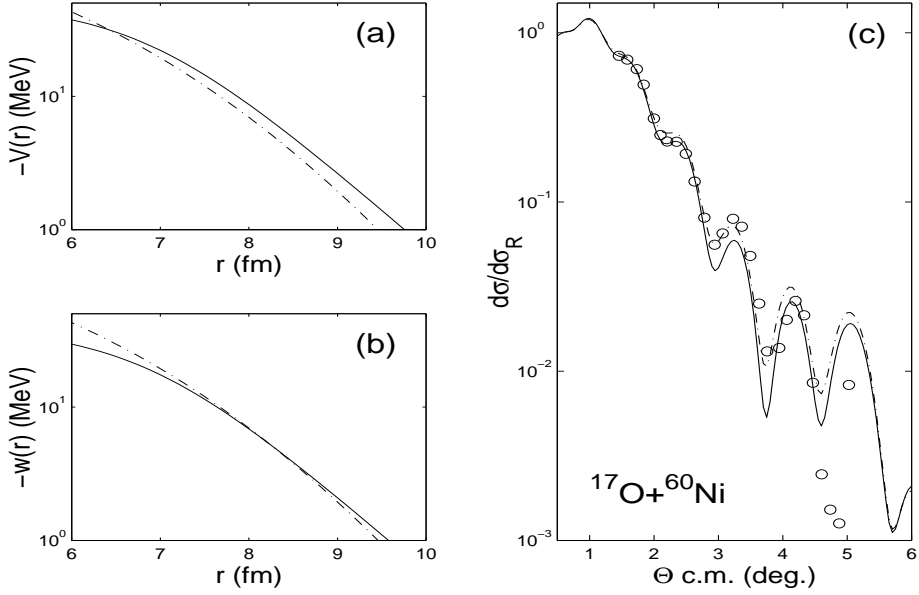,height=8cm,width=.8\linewidth}
\end{center}
\caption{The same as in Fig.1 but for scattering
$^{17}$O+$^{60}$Ni at 1435 MeV with optical potential $U_{opt}^C$
from Table 1B. Experimental points and the fitted WS-potential are
taken from \cite{Lig}.}
\end{figure}

\begin{figure}[t]
\begin{center}
\epsfig{file=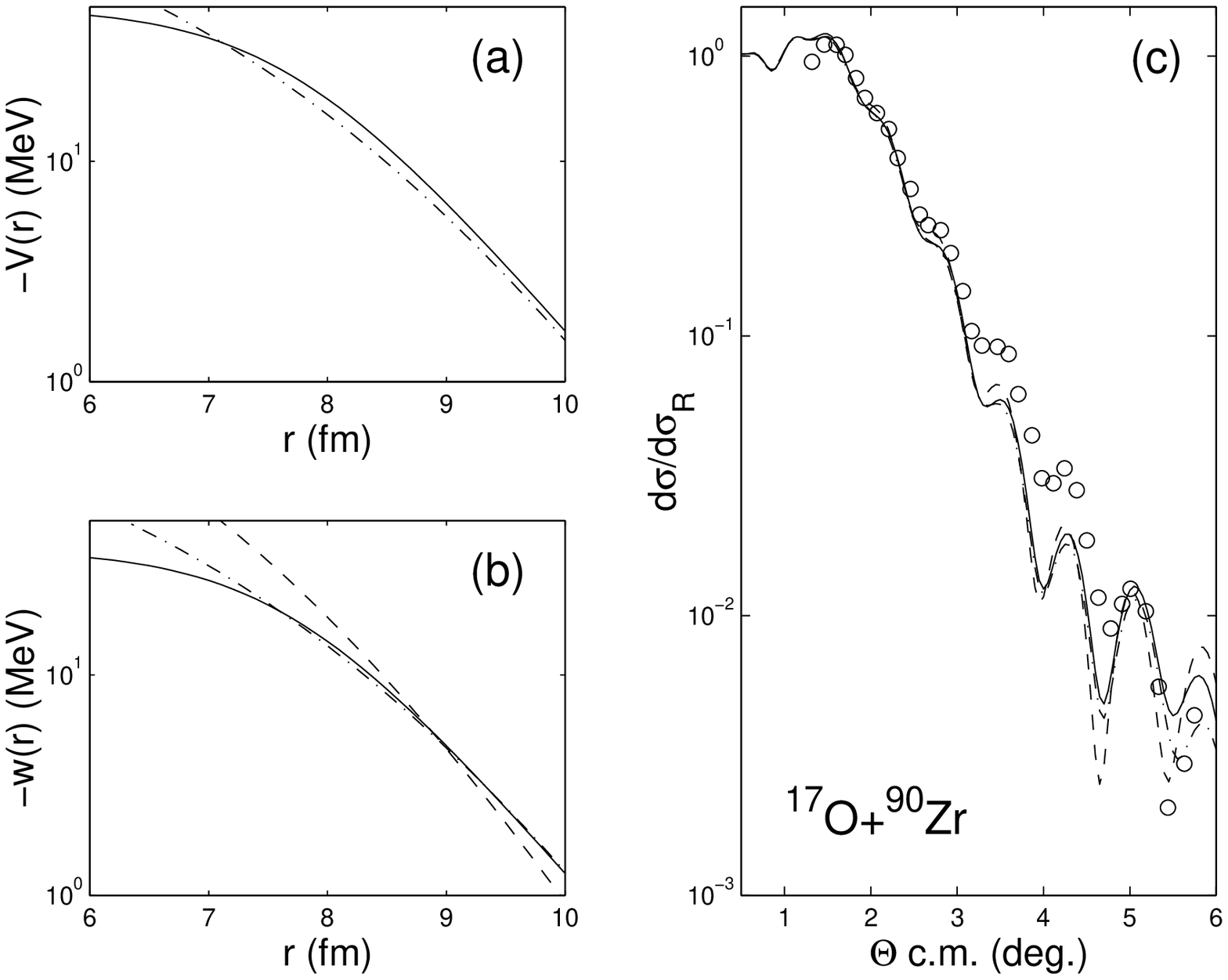,height=8cm,width=.8\linewidth}
\end{center}
\caption{The same as in Fig.4 but for scattering
$^{17}$O+$^{90}$Zr with optical potentials $U_{opt}^B$ and
$U_{opt}^C$.}
\end{figure}

\begin{figure}[t]
\begin{center}
\epsfig{file=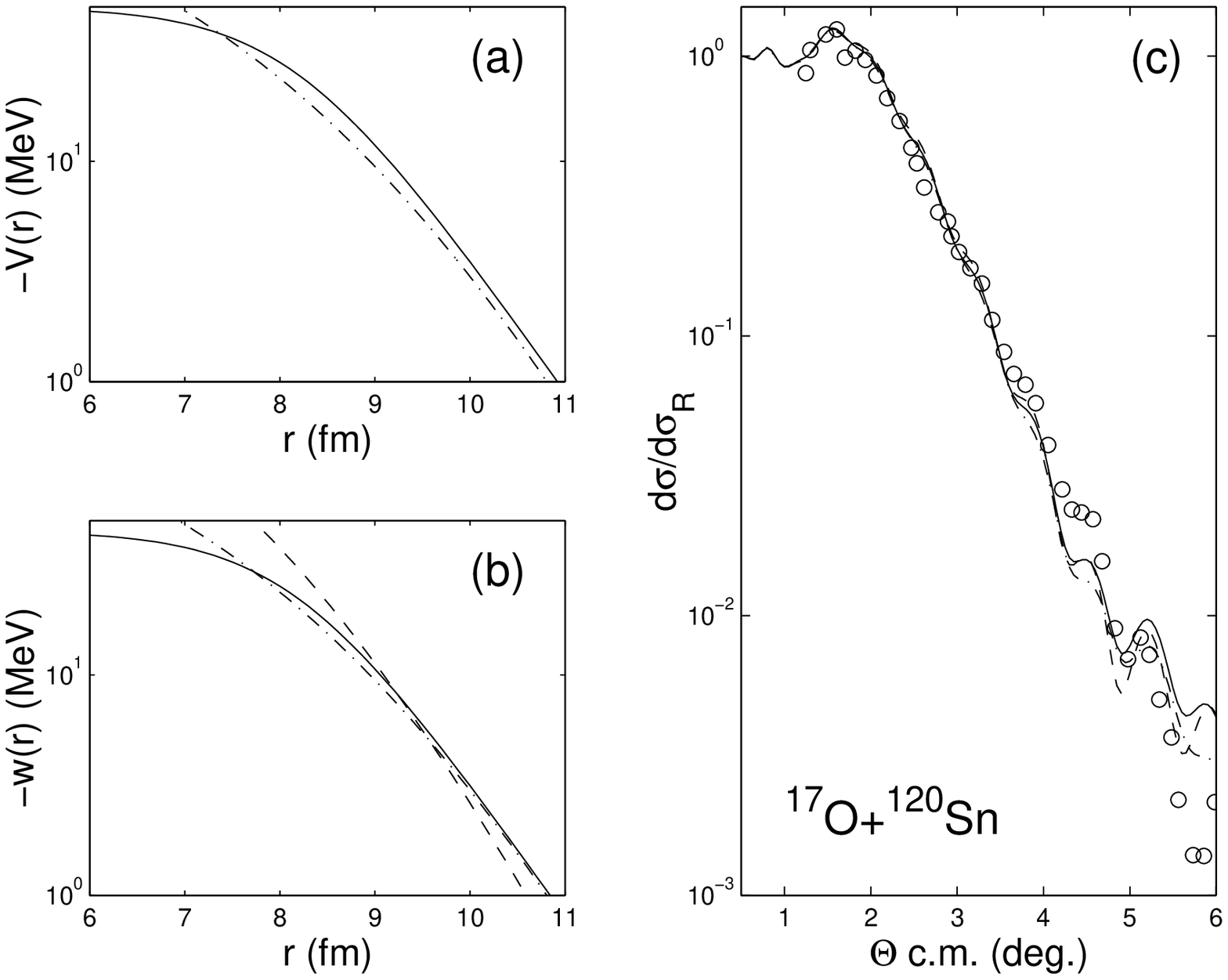,height=8cm,width=.8\linewidth}
\end{center}
\caption{The same as in Fig.4 but for scattering
$^{17}$O+$^{120}$Sn with optical potentials $U_{opt}^B$ and
$U_{opt}^C$.}
\end{figure}

\begin{figure}[t]
\begin{center}
\epsfig{file=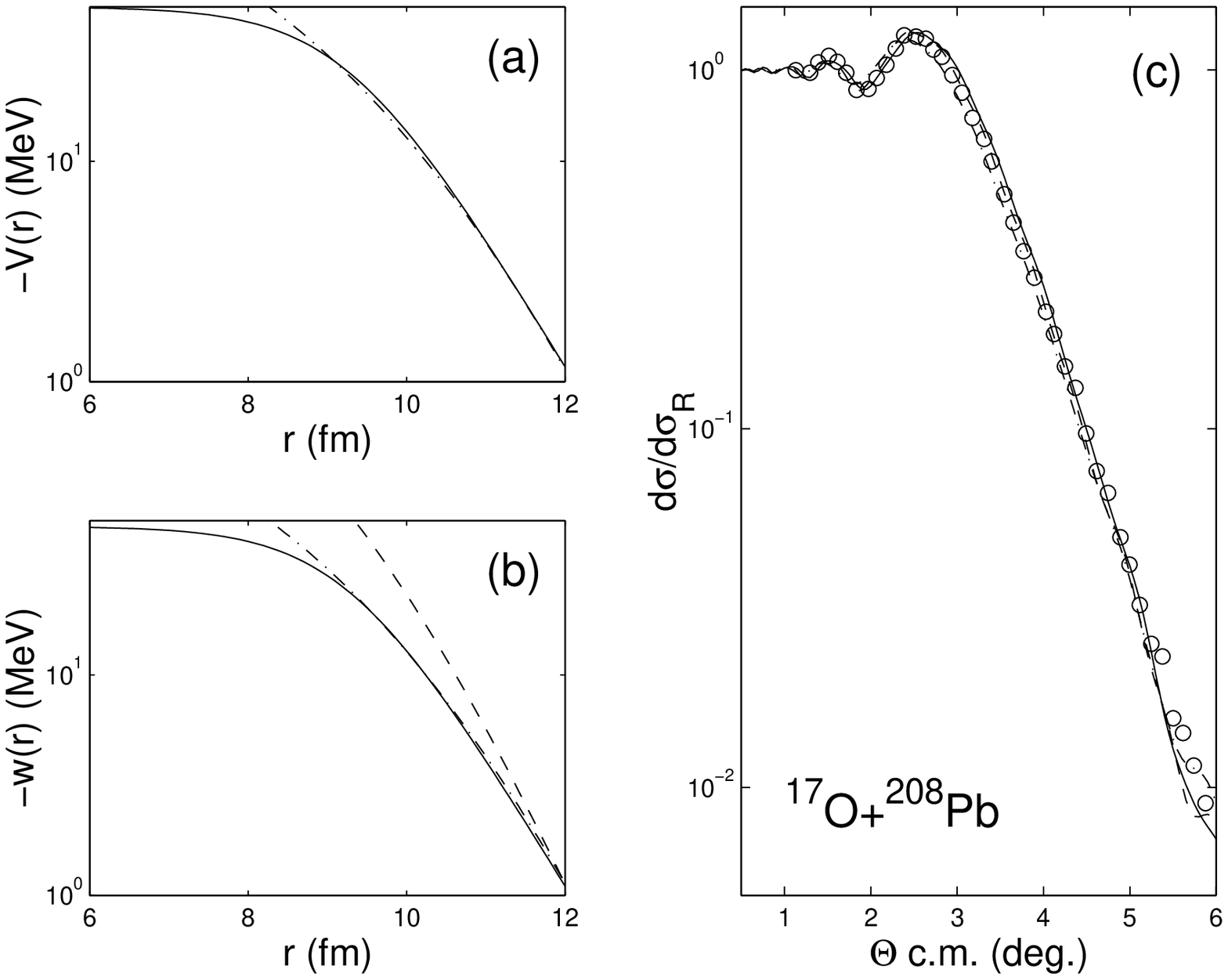,height=8cm,width=.8\linewidth}
\end{center}
\caption{The same as in Fig.4 but for scattering
$^{17}$O+$^{208}$Pb with optical potentials $U_{opt}^B$ and
$U_{opt}^C$. }
\end{figure}

\begin{figure}[t]
\begin{center}
\epsfig{file=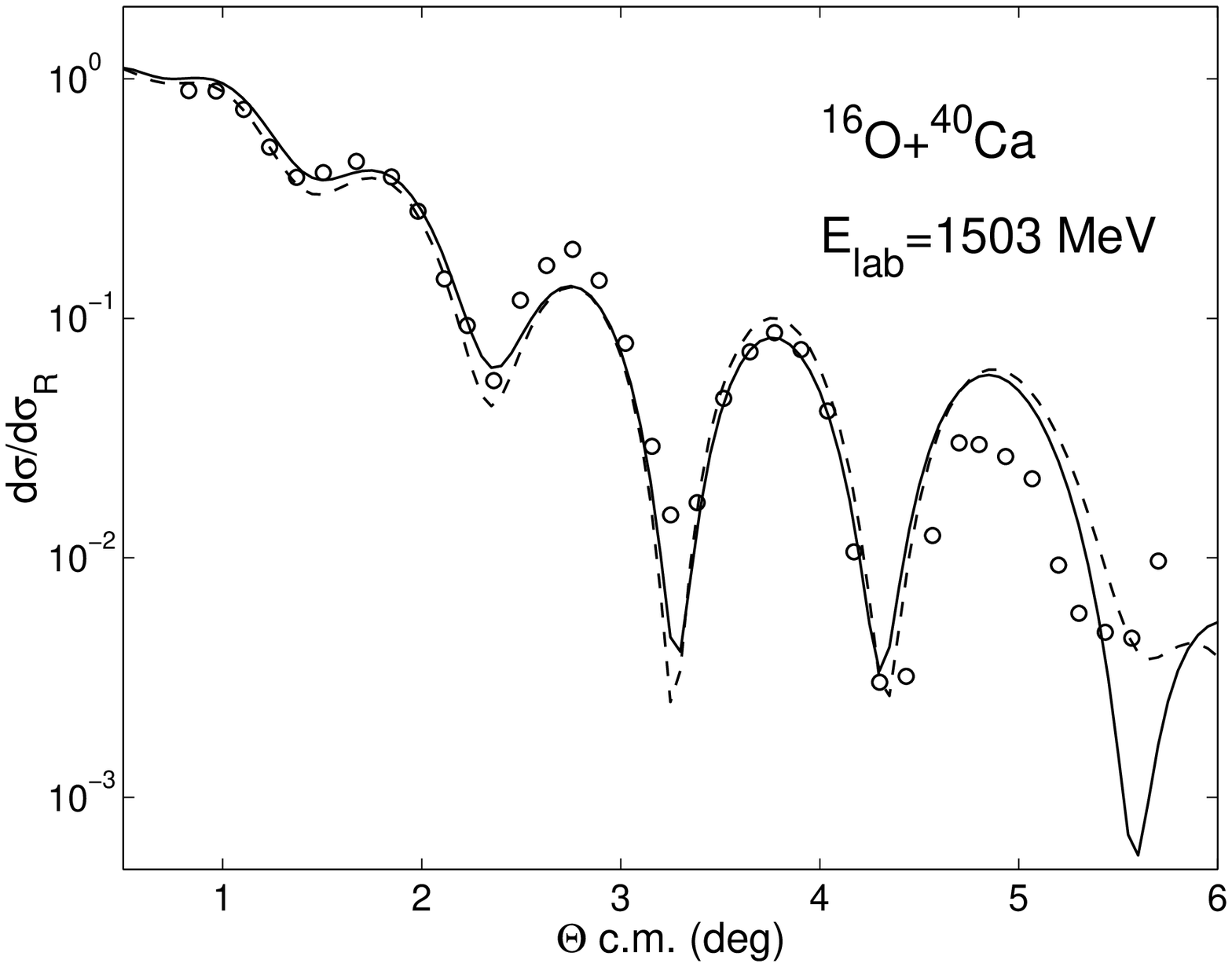,height=5.8cm,width=.55\linewidth}
\end{center}
\caption{The effect of relativization in case of scattering
$^{16}$O+$^{40}$Ca. Solid(dashed) lines show $d\sigma/d\sigma_R$~
with(without) relativization. The potential $U_{opt}^C$ is from
Table 1A.}
\end{figure}

The matter of fact is that, for nucleus-nucleus scattering, the surface region
of optical potentials plays a decisive role in predictions of differential
and total cross-sections. Concequantly, the usually ensured microscopic models
are substantiated namely in this outer region of the collision. Indeed, in a
preceding paper \cite{K1} a method was developed for the restoration of
nucleus-nucleus optical potentials derived on the basis of Glauber-Sitenko
microscopic scattering theory where, in the so called optical limit, the
microscopic phase was given in the form
\begin{equation}\label{eq2}
\Phi_N(b)={\bar{\sigma}_{NN}\over 2}(i+{\bar\alpha}_{NN})
\int d^2s_p\, d^2s_t~\ro_p(s_p)~\ro_t(s_t)~f_N(|\bf\xi=\bf b+{\bf s}_p-
{\bf s}_t|).
\end{equation}
Here $\ro_p(r)$ and $\ro_t(r)$ are the point nucleon density distributions
of the projectile and target nuclei, respectively, and  $\ro(s)=\int_{-\infty}
^\infty \ro(\sqrt{s^2+z^2})dz$ is the profile function of $\ro(r)$. Also,
the function $f_N(\xi)=(2\pi)^{-2}\int d^2q~\exp(-i{\bf q}{\bf\xi})
{\ti f}_N(q)$ is expressed through the form factor ${\ti f}_N(q)$ of the
NN-scattering amplitude, taken in the form ${\ti f}_N(q)=\exp(-q^2r_{N\,rms}
^2/6)$ with $r_{N\,rms}$, the NN-interaction ${rms}$ radius. Here $\bar\sigma
_{NN}$ is the total cross section of the NN-scattering while $\bar\alpha_{NN}$
is the ratio of the real-to-imaginary part of the forward NN-scattering
amplitude, and both of these quantities depend on energy. We denote that
the "bar" means averaging on isotopic spins of colliding nuclei. In \cite{K1},
this phase (\ref{eq2}) was compared with another phenomenological one defined
through the optical potential $U(r)=V(r)\,+\,iW(r)$ as follows,
\begin{equation}\label{eq3}
\Phi(b)=-\frac 1 {\hbar v}\, \int_{-\infty}^\infty U\el(\sqrt{b^2+z^2}\er)dz,
\end{equation}
where $v$ is the relative motion velocity. An analytic expression was used
for the phase $\Phi(b)$ of (\ref{eq3}), obtained in \cite{LZ} for the
symmetrized Woods-Saxon (SWS) potential, which is the most realistic
phenomenological potential often applied in many calculations. The parameters
of the SWS-potential were adjusted such that to fit the shape of the
phenomenological phase (\ref{eq3}) to the microscopic one (\ref{eq2}) in
the outer region of space $b\sim R_p\,+ \,R_t$. As a result of this procedure
it was obtained a set of SWS-potentials which coincide in their tails but
have different interiors, and all of them were in a reasonable agreement
with elastic scattering differential cross-sections at small angles. Although
this method gave surface-equivalent realistic WS-type potentials which means
the exclusion of ambiguities in the peripheral region of the interaction,
it puts us in face of the traditional old standing ambiguity problem of the
optical potentials especially in their internal region.

In such situation, we intend in this work to suggest another
approach to restore an optical potential. Towards this aim we
believe that the use of microscopic potential models is more
reliable in search of a realistic optical potential than fitting a
phenomenological one. As a first candidate in this search we
suggest to use unambiguous potential that corresponds to the HEA
microscopic phase (\ref{eq2}). This potential has been obtained
independently in \cite{S1}, by applying the inverse Fourier
transform to the HEA-phase (\ref{eq2}), and in \cite{lzl}, by
substituting the standard expression for the direct DF-potential
in the definition of the phase (\ref{eq3}). As a result,  the
so-called HEA-optical potential is as follows:
\begin{equation}\label{eq4}
U^H_{opt}(r)\,=\, V^H(r)\,+\,iW^H(r),
\end{equation}
where
\begin{equation}\label{eq5}
V^H(r)=-{2E\over k(2\pi)^2}{\bar\sigma}_{NN}{\bar\alpha}_{NN}
\int dq~q^2j_0(qr){\ti\ro}_p(q){\ti\ro}_t(q){\ti f}_N(q),
\end{equation}
\begin{equation}\label{eq6}
W^H(r)=-{2E\over k(2\pi)^2}{\bar\sigma}_{NN}
\int dq~q^2j_0(qr){\ti\ro}_p(q){\ti\ro}_t(q){\ti f}_N(q).
\end{equation}
Here ${\ti\ro}_{p(t)}(q)$ are form factors of the corresponding point
densities $\ro_{p(t)}(r)$ of the projectile and target nuclei, where
the latter functions can be obtained by unfolding the nuclear densities
$\rr_{p(t)}(r)$ (see, e.g., \cite{lzs}), which are usually given in tabulated
forms. Thus, the suggested model is free from parameters when calculating
the real $V^H$ and the imaginary $W^H$ parts of the potential. The important
and novel point of this method is that it provides to calculate the imaginary
part of the potential (\ref{eq6}) in a microscopic way. We remind, that in
the standard semi-microscopic model one estimates only the real part of the
potential using DF-procedure, while the imaginary part is usually taken in
a phenomenological WS-form with three or sometimes more fitted parameters.
In the present work, in addition to the HEA-potential, we also apply a
DF-procedure to estimate the real part of the optical potential, which
includes both the direct and exchange terms (see, e.g., \cite{{Knyaz},{KS}}):
\begin{equation}\label{eq7}
V^{DF}=V^D~+~V^{EX}
\end{equation}
where
$$
\qquad V^D(r) = \int d^3 r_p d^3 r_t \, \rho_p({\bf r}_p)\,
\rho_t({\bf r}_t)\, v_{NN}^D({\bf r}_{pt}), \quad
{\bf r}_{pt}={\bf r}+{\bf r}_t-{\bf r}_p,
\eqno (7a)
$$
$$
\qquad V^{EX}(r) = \int d^3 r_p d^3 r_t \, \rho_p({\bf r}_p, {\bf r}_p+
{\bf r}_{pt})\,\rho_t({\bf r}_t, {\bf r}_t-{\bf r}_{pt})\times \hspace*{4cm}
$$
$$
\hspace*{8cm}  v^{EX}_{NN}({\bf r}_{pt})\,\exp\el[{i{\bf K}(r)
{\bf r}_{pt}\over M}\er].
\eqno (7b)
$$
The dependence on energy in the potential comes from the local relative
momentum motion defined as $K(r)\simeq\{2Mm/\hbar^2[E-V_N(r)- V_C(r)]\}^
{1/2}$ where $Mm=A_pA_tm/(A_p+A_t)$ is the reduced mass, E is the relative
energy in the center-of-mass frame, and $V_C(r)$, the responsible part of
the interaction due to the Coulomb potential. We adopt here an energy- and
density-dependent version for the effective interaction as given in
\cite{KS} where the effective interaction $v_{NN}$ is expressed in the
form of M3Y force multiplied by the factor $F(\rho)=C[1+\alpha\exp(-\beta\rho)
-\gamma\rho]$ which depends on the densities $\rho= \rho_p + \rho_t$, and
also the additional factor $(1-0.003\,E/A_p)$ is introduced to correct
the dependence upon the incident laboratory energy per nucleon.

The comparison between (\ref{eq5}) and (\ref{eq7}) ensures that
the HEA real part $V^H$ of the optical potential corresponds only
to the direct part $V^D$ of the full potential while the
$V^{DF}$-real potential consists of two terms, direct and exchange
ones, where the latter has a nonlocal nature and arises from the
anti-symmetrization between two colliding nuclei, and it accounts
for the Pauli-blocking and the so-called knock-on exchange
nonlocality. Thus we have two microscopic types of the real
potentials $V^{DF}$ and $V^H$, and one for the imaginary part
$W^H$. The HEA-potentials have slightly different slopes in their
asymptotics as compared to the DF-potential. In principle, the
real and imaginary parts of optical potentials have different
physical nature. The first one, as its origin, has the
one-particle densities while the second one can get the additional
contributions, coming from excitations of collective states and the
nucleons removal reactions. Besides, one should bear in mind that
at high energies, the peripheral region of the nucleus-nucleus
interaction plays the essential role, while the exchange effects
reveal themselves mainly in the internal region. At the same time,
we pay attention to the result given in \cite{W} that at high
energies the nucleons removal reactions mostly contribute to the
absorption part of the optical potential while the excitation
channels are suppressed. Therefore, one-particle densities take
part in equal footing in the formation of both the real and imaginary
potentials. Thus, considering not high but intermediate energies
of collisions at about 100 MeV/nucleon one can utilize both the
shapes HEA- and DF-patterns for composing total microscopic
potentials. As a result, we shall test three types of optical
potentials, each have only two parameters $N_{r}$ and $N_{im}$,
namely:
\begin{equation}\label{eq8}
U^A_{\scr{opt}} ~ = ~ N^A_{r}\,V^H ~ + ~ iN^A_{im}\,W^H
\end{equation}
\begin{equation}\label{eq9}
U^B_{\scr{opt}} ~ = ~ N^B_{r}\,V^{DF} ~ + ~ iN^B_{im}\,W^H
\end{equation}
\begin{equation}\label{eq10}
U^C_{\scr{opt}} ~ = ~ N^C_{r}\,V^{DF} ~ + ~ iN^C_{im}\,V^{DF}
\end{equation}
Usually, in heavy-ion scattering at comparably high energies, the
potential tails determine the pattern of the elastic differential
cross-sections because of the strong absorption happened at
shorter distances. Then, roughly speaking one needs only four
parameters to describe the positions and the slope parameters of
these tails. In our case we use the microscopic models for both
the real and imaginary patterns of the optical potentials given by
Eqs.(\ref{eq8})-(\ref{eq10}), where by the fitting of only two
parameters $N_r$ and $N_{im}$ we can, in fact, change the strength
and shift of the potential tails in the surface region. In
practice, the fit of phenomenological potentials at $E\sim 100$
Mev/nucleon shows that the range from $R_{in}$ to $\infty$
determines the main pattern of the differential cross-sections,
and $R_{in}$ is the radius where $V(R_{in})=-50$ Mev. So, below in
Figures we show potentials only in this region of their
displaying.

\section {Results, Discussion, and Conclusions}

We calculate the ratio of the elastic differential cross-sections $d\sigma/
d\Omega= |f(q)|^2$ to the Rutherford cross-section
\begin{equation}\label{eq11}
{d\sigma_R\over d\Omega}\,= \, \el ({Z_pZ_te^2\over \hbar v}\er )^2\,
{1\over 4k^2}\, {1\over\sin^4(\vartheta/2)}.
\end{equation}
For this purpose we apply the expression for the HEA-scattering amplitude
\begin{equation}\label{eq12}
f(q) = ik \int_0^\infty db b\, J_0 (qb)\Bigl [1-{\dis e}^
{\dis i\Phi_N(b)+i\Phi_C(b)}\Bigr ].
\end{equation}
\\
which is valid at $E\gg |U|$ and for small scattering angles
$\vartheta < \sqrt{2/kR}$ where $R$ is the nucleus-nucleus interacting radius,
say, $R\sim {R_p+R_t}$. Here $q=2k\sin(\vartheta/2)$ is the momentum transfer.
The Coulomb phase $\Phi_C(b)$ is taken in an analytic form for the uniformly
charged spherical density distribution. The nuclear phase $\Phi_N(b)$ is
calculated with a help of the optical potentials (\ref{eq8})-(\ref{eq10}),
using the microscopic HEA- and DF-models. The trajectory distortion
in the Coulomb field is taken into account by changing the impact parameter
$b$ by $b_c=\bar a+\sqrt{{\bar a}^2+b^2}$ in all functions of the integrand
of (\ref{eq12}) with the exception of $\Phi_C(b)$; here $b_c$ is the
distance of closest approach in a Coulomb field, where $\bar a=Z_pZ_te^2/2E_
{c.m.}$. Details of calculations of (\ref{eq12}) one can find in \cite{LZ2}.
In addition, in calculations, we take into account the relativistic kinematics
by substituting the respective expressions of velocity $v$ and the c.m.
momentum $k$ in (\ref{eq3}), (\ref{eq11}) and (\ref{eq12}) as follows:
\begin{equation}\label{eq13}
\hbar v\,= \,197.327\,{\sqrt{E_l(E_l+2A_pm)}\over E_l+A_pm} \quad
(in ~~MeV\,fm),
\end{equation}
\begin{equation}\label{eq14}
k\,=\,{1\over 197.327}\,\frac{A_t\sqrt{E_l(E_l+2A_pm)}}
{\sqrt{(A_p+A_t)^2\,+\,2A_tE_l/m}} \quad (in~~fm^{-1}),
\end{equation}
\\
where $E_l$ (in MeV) is the kinetic energy of the projectile nucleus in
laboratory system, and $m$=931.494 (in MeV) is the unified atomic mass unit.

Below we present our calculations of the cross-section
$d\sigma/d\sigma_R$ for scattering of $^{16}$O on the targets
$^{40}$Ca,~$^{90}$Zr, and $^{208}$Pb at incident energy $E_l$=1503
MeV, and $^{17}$O on $^{60}$Ni, $^{90}$Zr, $^{90}$Sn, and
$^{208}$Pb at $E_l$=1435 MeV, and compare these calculations with
the corresponding experimental data from Refs. \cite{Ro} and
\cite{Lig}, respectively. The pattern potentials $V^H,~W^H$, ~and
$V^{DF}$ were computed with the help of (\ref{eq4})-(\ref{eq7})
using the point density distribution functions $\ro(r)$ for $^{16}$O
and respective target-nuclei from \cite{lzs}, and for the corresponding
nuclei in collisions of $^{17}$O - from \cite{PP} and \cite{FaS}. Also,
parameterization of $\bar\sigma_{NN}$ and $\bar\alpha_{NN}$ are taken
from \cite{CG} and \cite{S2} while the effective $v_{NN}$-forces of
the type CDM3Y6 are obtained from \cite{KSO}. The normalizing coefficients
$N_r$ and $N_{im}$ in (\ref{eq8})-(\ref{eq10}) were fitted for each couple
of colliding nuclei and presented in Tables 1A and 1B.

In Figs.1-7, panels (a) and (b) show the real and imaginary parts of the
optical potentials $U_{opt}^A$, $U_{opt}^B$, and $U_{opt}^C$ calculated
by using the microscopic models HEA and DF as patterns. Dashed curves
represent the potentials with patterns $V^H$ and $W^H$, while those with
$V^{DF}$ are shown by dash-dotted curves. The phenomenological Woods-Saxon
(WS) potentials, are shown by solid lines. The ratios of the respective
elastic to Rutherford differential cross-sections are presented in panels
(c) of Figs.1-7, where dashed curves show the HEA-calculations
with the potentials $U_{opt}^A$ or $U_{opt}^B$, dashed-dotted lines -- with
the potentials $U_{opt}^C$, and solid curves -- with the fitted
WS-potentials; open circles are the experimental data.

One can see that the slopes of the calculated and the fitted
potentials in the outer region have a coincidence to each others.
The differential cross-sections fall down by an exponential low
beyond the Coulomb rainbow angle, and have an acceptable agreement
with the experimental data. As to applicability of the
HEA-calculations, we can refer to the sufficient agreements with
the experimental data of the HEA cross-sections for the
WS-potentials (solid curves). On the other hand, these potentials
were obtained by fitting to the same data given in \cite{Ro} and
\cite{Lig} not by the HEA-calculations but with the help of
numerical solutions of the Schroedinger equation. Indeed, this
agreement takes place at angles $\vartheta < 5.5^\circ $ where the
HEA is valid by definition. In Tables 1A and 1B the fitted
normalizing factors $N_r$ and $N_{im}$ of both the real and
imaginary parts of the different microscopic optical potentials
are demonstrated. In addition, we demonstrate in Fig.8 the
relativistic effect on the differential elastic scattering
cross-section of $^{16}O$+$^{40}$Ca at $E_{l}$=1503 MeV, when one
uses the relativistic formulae (\ref{eq13}),(\ref{eq14}) for $k$
and $v$ in (\ref{eq3}),(\ref{eq11}), (\ref{eq12}). Although this
effect is seen not to be large at this energy, the calculated
cross-section $d\sigma/d \sigma_R$ is in favor of its improvement
when compared with its experimental counterpart.

\vspace{0.2cm}
{\samepage \hspace*{0.3cm} { {\bf Table 1A.}~~{\it Optical
potentials for the $^{16}$O heavy-ion scattering on different nuclei}}

\begin{center}
\begin{tabular}{|c|c|c|c|}
\hline \hline
$ U $             
&${^{16}O}+{^{40}Ca}$&${^{16}O}+{^{90}Zr}$&${^{16}O}+{^{208}Pb}$ \\
\hline \hline
$U^A_{opt}$& ---    & $1.13V^{H}+iW^{H}$& --- \\

$U^B_{opt}$&$V^{DF}+i1.32W^H$&---&$V^{DF}+iW^H$ \\
$U^C_{opt}$&$V^{DF}+i0.88V^{DF}$&---&$V^{DF}+i0.6V^{DF}$  \\
\hline
\end{tabular}
\end{center}
}

\vspace{.5cm}
{\samepage \hspace*{0.3cm} { {\bf Table 1B.}~~{\it Optical
potentials for the $^{17}$O heavy-ion scattering on different nuclei}}

\begin{center}
\begin{tabular}{|c|c|c|c|c|}
\hline \hline
 U             
&${^{17}O}+{^{60}Ni}$&${^{17}O}+{^{90}Zr}$&${^{17}O}+{^{120}Sn}$&${^{17}O}+{^{208}Pb}$ \\
\hline \hline
$U^A_{opt}$& ---    & ---  & --- & --- \\

$U^B_{opt}$& --- &$0.6V^{DF}+i0.9W^H$&$0.5V^{DF}+i0.9W^H$&$0.5V^{DF}+i1.3W^H$ \\
$U^C_{opt}$&$0.6V^{DF}+i0.6V^{DF}$&$0.6V^{DF}+i0.5V^{DF}$&$0.5V^{DF}+i0.5V^{DF}$
&$0.5V^{DF}+i0.8V^{DF}$ \\
\hline
\end{tabular}
\end{center}
}

\vspace{0.3cm} Our main conclusion, although we did not intend to
achieve a perfect fit as usually experimentalists do, is that the
presented idea proves itself to utilize the microscopic models as
patterns for further fit with the experimental data. In addition,
this method introduce only two adjusted normalizing free
parameters instead of, at least, twice that number of parameters
required in case of use the phenomenological WS-optical potential.
Moreover, at high energy interactions, one can be confident
to claim that the results of the calculations done by using the
microscopic potentials show that in the outer region of the
interactions a true prediction and behavior of these potentials
can be gained in the very sensitive domain of the heavy-ion
scattering. \vspace*{.2cm}
\begin{center}
{\large ACKNOWLEDGMENTS}
\end{center}

The co-authors V.K.L. and B.S. are grateful to the Infeld-Bogoliubov
Program for support of this work. One of us E.V.Z. thanks the Russian
Foundation Basic Research (project 03-01-00657) for support. Also deep
gratitude from K.M. Hanna to the authorities of AEA-Egypt and JINR for
their support.


\end{document}